\documentclass[a4paper]{jpconf}
\usepackage{graphicx}

\usepackage{lineno}

\begin{document}
\title{Awkward to RDataFrame and back}

\author{Ianna Osborne, Jim Pivarski}

\address{Princeton University, Princeton, NJ 08544, USA}

\ead {ianna.osborne@cern.ch}

\begin{abstract}
Awkward Arrays~\cite{awkward-ref} and RDataFrame~\cite{rdf-ref} provide two very different ways of performing calculations at scale. By adding the ability to zero-copy convert between them, users get the best of both. It gives users a better flexibility in mixing different packages and languages in their analysis.
In Awkward Array version 2, the {\tt ak.to\_rdataframe} function presents a view of an Awkward Array as an RDataFrame source. This view is generated on demand and the data are not copied. The column readers are generated based on the run-time type of the views. The readers are passed to a generated source derived from ROOT::RDF::RDataSource. 
The {\tt ak.from\_rdataframe} function converts the selected columns as native Awkward Arrays. 
The details of the implementation exploiting JIT techniques are discussed. The examples of analysis of data stored in Awkward Arrays via a high-level interface of an RDataFrame are presented. 
A few examples of the column definition, applying user-defined filters written in C++, and plotting or extracting the columnar data as Awkward Arrays are shown. Current limitations and future plans are discussed.

\end{abstract}

\section{Introduction}

Awkward Array is a library for nested, variable-sized data, including arbitrary-length lists, records, mixed types, and missing data, using NumPy-like~\cite{numpy-ref} idioms. 

In a typical Awkward user analysis workflow the access to columnar data is provided by ServiceX~\cite{service-x-ref} and Uproot~\cite{uproot-ref}. The data are presented to the user as Awkward Arrays. The user performs the event selection, applies the systematic uncertainties, produces histograms, builds a statistical model, likelihoods, statistical analysis, fit results and diagnostics. The user has access to a wide range of Python ecosystem packages - especially the Scikit-HEP packages~\cite{sci-kit-hep-ref} - to perform each task.

In contrast, the RDataFrame is a declarative, parallel framework for data analysis and manipulation. It reads from a columnar data format via a data source. It applies transformations to the data - that is, selects rows, defines new columns - and produces results. The results can be data reductions like histograms, new ROOT files, or any other user-defined object or side effect.

\section{Awkward Array in Python eco-system}

\begin{figure}[htp]
    \centering
    \includegraphics[width=12cm]{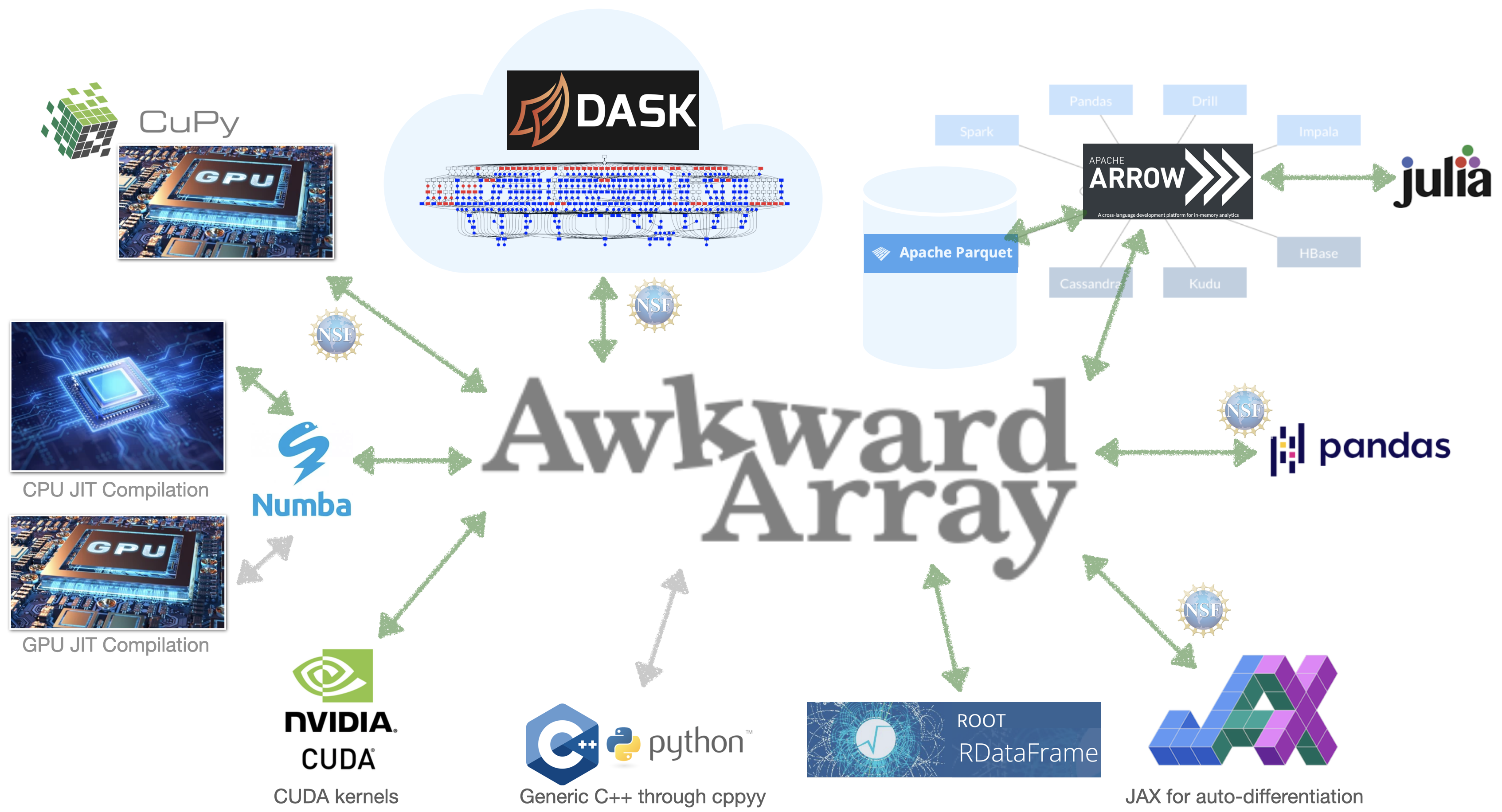}
    \caption{Awkward Array interoperability with other projects}
    \label{fig:AwkwardArray_and_other_projects}
\end{figure}

Awkward Arrays and RDataFrame are two very different ways of performing large-scale calculations. The {\tt Awkward-RDataFrame} bridge provides users with more flexibility in mixing different packages and languages in their analyses if desired (see Fig.~\ref{fig:AwkwardArray_and_other_projects}). There are numerous benefits of combining both Python and C++. The users can mix analyses using Awkward Arrays, Numba~\cite{numba-ref}, and ROOT C++ in memory, without saving to disk and without leaving their environment.

On the one hand, the users who do their analysis entirely in Python eco-system can benefit from faster execution using ROOT C++ functions, or pure C++, in an otherwise Awkward analysis at full speed. The performance cost of converting Awkward Arrays into RDataFrame is negligible, since it is a zero-copy view, and the conversion of RDataFrame lists into Awkward Arrays is discussed below in section~\ref{From RDataFrame to Awkward Array}.

On the other hand, those who prefer the C++, ROOT, and RDataFrame, have an ability to convert their data into Awkward Arrays. This conversion opens many paths to follow.

\section{From Awkward Array to RDataFrame}

\begin{figure}[htp]
    \centering
    \includegraphics[width=3in]{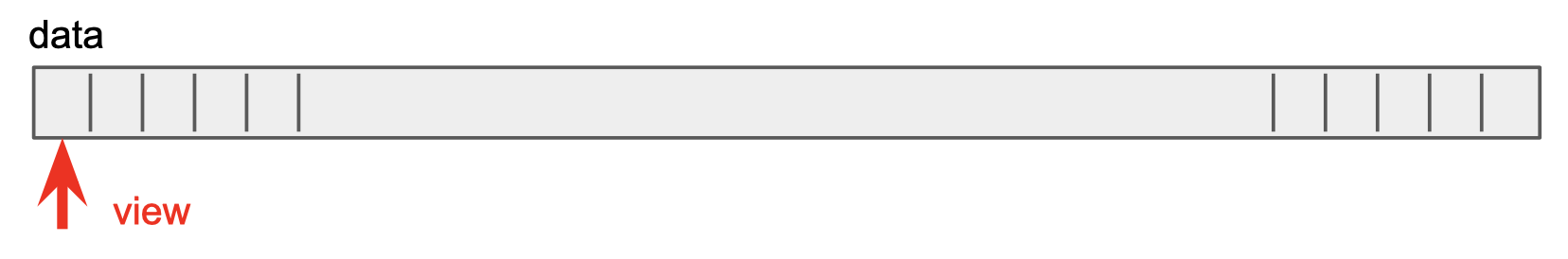}
    \caption{Awkward Array View can be thought as a cursor}
    \label{fig:AwkwardArray_ArrayView_cursor}
\end{figure}

The Awkward-style {\tt ak.to\_rdataframe} function~\cite{ak_to_rdataframe} presents a view of an Awkward Array as an RDataFrame source. The view is a lightweight 40-byte C++ object dynamically allocated on the stack. The generated RDataSource takes pointers into the original array data via this view. This view is generated on demand, the data are not copied - see figure~\ref{fig:AwkwardArray_ArrayView_cursor}. The view and the array data are accessible for as long as the returned by this function RDataFrame has not completed its execution and is currently in a running state.

This work takes advantage of an existing feature, in which Awkward Arrays can be iterated over in Numba-compiled functions, again using zero-copy views. The Numba implementation for C++ is being reused here: there is no performance difference. The column readers are generated based on the run-time type of the views. The readers are passed to a generated source derived from RDataSource.

The {\tt ak.to\_rdataframe} function takes a dict as its argument: each key defines a column name in the RDataFrame.
The equal length arrays are given as values to the dictionary keys. The arrays data are not copied, but there is a small overhead of generating an Awkward RDataSource C++ code.

The following example describes three typical and distinct Awkward Arrays: an array of records, a flat array, and a list of arrays. Passing these arrays to RDataFrame creates three different columns with three different types.

The column {\tt "x"} will maintain its awkward type, while the column {\tt "y"} will become a column of integers, and the column {\tt z} - a variable length array: each array is a container ({\tt RVec}) of {\tt double} values. The complete tutorial~\cite{awkward-rdf-pyhep2022-ref} including a toy analysis on CMS open data~\cite{cms-data} has been presented at PyHEP2022 workshop.

\begin{verbatim}
    import awkward as ak
    import ROOT

    array_x = ak.Array([
        {"x": [1.1, 1.2, 1.3]}, 
        {"x": [2.1, 2.2]}, 
        {"x": [3.1]}, 
        {"x": [4.1, 4.2, 4.3, 4.4]}, 
        {"x": [5.1]},])
    array_y = ak.Array([1, 2, 3, 4, 5])
    array_z = ak.Array([[1.1], [2.1, 2.3, 2.4], [3.1], [4.1, 4.2, 4.3], [5.1]])

    assert len(array_x) == len(array_y) == len(array_z)

    df = ak.to_rdataframe({"x": array_x, "y": array_y, "z": array_z})
\end{verbatim}

This {\tt ak.to\_rdataframe} operation does not execute the RDataFrame event loop. The Awkward {\tt RDataSource} is interpreted as Custom and the columns are its Datasets:

\begin{verbatim}
    Dataframe from datasource Custom Datasource

    Property                Value
    --------                -----
    Columns in total            4
    Columns from defines        1
    Event loops run             0
    Processing slots            1

    Column          Type                            Origin
    ------          ----                            ------
    x               awkward::Record_DZ9qK2aXbBA     Dataset
    y               int64_t                         Dataset
    z               ROOT::VecOps::RVec<double>      Dataset
\end{verbatim}

The RDataFrame column type is a string that corresponds to a C++ data type. The Awkward data types are defined in an {\tt "awkward"} C++ namespace. Here, for example, the {\tt "x"} column contains an Awkward Array with a made-up type: {\tt awkward::Record\_DZ9qK2aXbBA}.
Awkward Arrays are dynamically typed and, in a C++ context, the type name is dynamically generated, and the name contains a hash of its contents to ensure uniqueness. In practice, there is no need to know the type. The C++ user code should use a placeholder type specifier {\tt auto} - the type of the variable that is being declared will be automatically deduced from its initialiser.

The RDataFrame framework can perform all usual operations on Awkward data with one exception, the Snapshot operation. Presently, the Awkward type columns are not saved to a ROOT file. Scheduling an operation does not execute the event loop. For example, here is a filtering operation on all data where the column {\tt "y"} values are greater than {\tt 2}:

\begin{verbatim}
    df = df.Filter("y % 2 == 0")
\end{verbatim}

The filtered Awkward Array internal layout - a RecordArray data: its content NumpyArray - is not copied, it is indexed. It is wrapped in an IndexedArray - because of the filter selection. The other two columns data are copied. The same operation on Awkward arrays in Python produces the same result, the array of the same type, but a different internal layout - because Awkward arrays are immutable.

\begin{verbatim}
    array_xyz = ak.Array({"x": array_x, "y": array_y, "z": array_z})
    filtered_array = array_xyz[array_xyz["y"] % 2 == 0]
\end{verbatim}

\section{From RDataFrame to Awkward Array}
\label{From RDataFrame to Awkward Array}

\begin{figure}[htp]
    \centering
    \includegraphics[width=12cm]{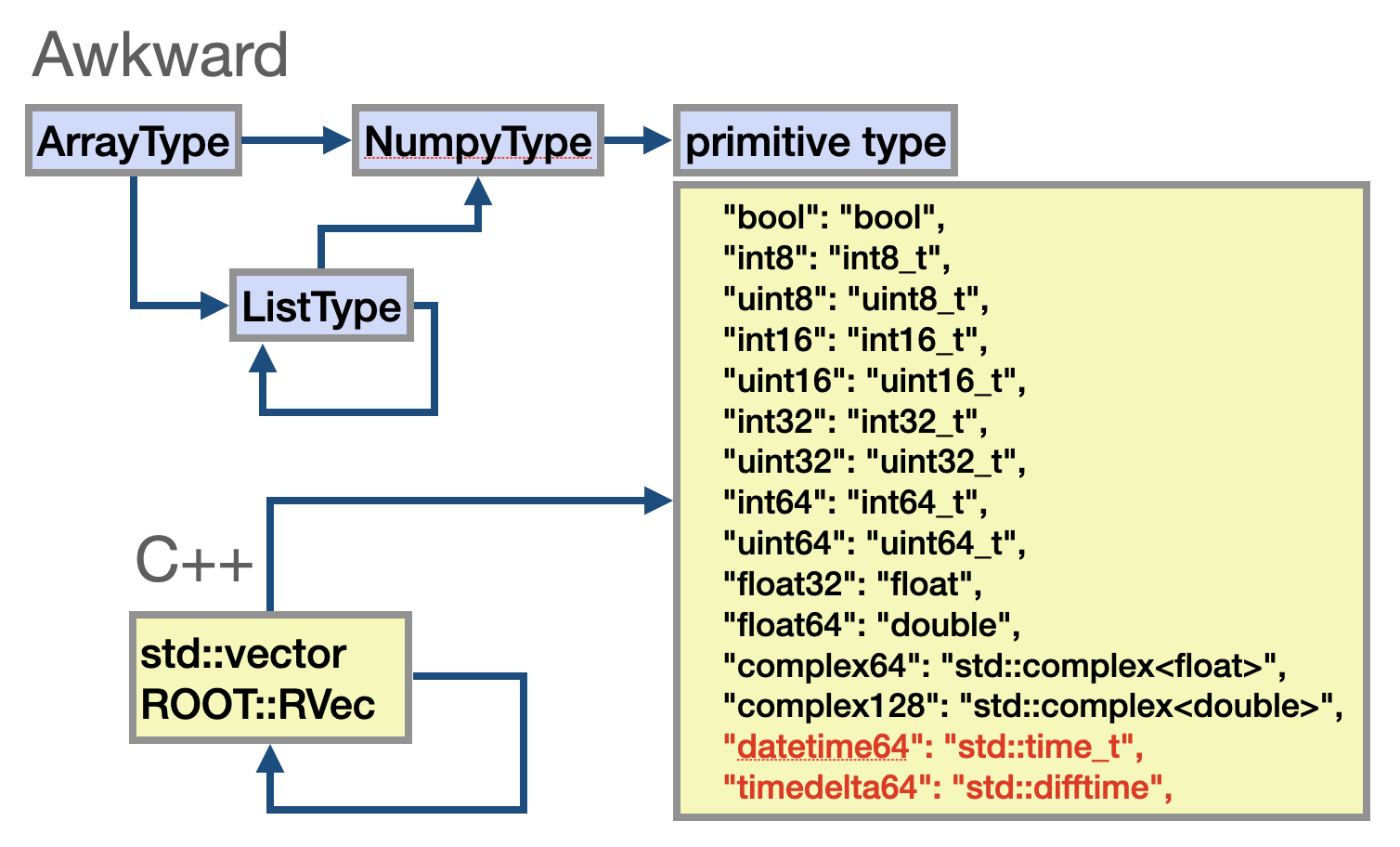}
    \caption{Supported Awkward Array and RDataFrame types. The last two lines under "primitive type" are red: they're not supported. An equivalent C++ type from the {\tt std} C-style date and time utilities library hasn't been mapped to these Awkward types.}
    \label{fig:AwkwardArray_RDataFrame_types}
\end{figure}

The {\tt ak.from\_rdataframe} function~\cite{ak_from_rdataframe} converts selected columns to native Awkward Arrays. The function takes a string or a tuple of strings that are the RDataFrame column names and recognises the following column data types (see figure~\ref{fig:AwkwardArray_RDataFrame_types}.):

\begin{itemize}
    \item Primitive types: $integer$, $float$, $double$, $std::complex<double>$, etc.
    \item Lists of primitive types and the arbitrary depth nested lists of primitive types: $std::vector<double>$, $RVec<int>$, etc.
    \item Awkward types: the run-time generated types derived from $awkward::ArrayView$ or $awkward::RecordView$. There is no copy required because Awkward Arrays are immutable - a reference to the original input array is passed to the output.
\end{itemize}

The RDataFrame event loop is triggered once to retrieve all selected columns.

\begin{verbatim}
    out = ak.from_rdataframe(df, columns=("x", "y", "z",), )
\end{verbatim}

Both the C++ templated header-only Awkward-cpp implementation and the dynamically generated C++ code are used to extract the column types. This approach simplifies the JIT-compilation in ROOT. The array Python string description is constructed from the C++ data types. Both the dynamically generated from the Python string Layout builder~\cite{layout-builder} and the generated functions are needed to process the column data and are compiled with Cling~\cite{cling-ref}.

\section{Summary}
Awkward Arrays and RDataFrame provide two very different ways of performing large scale calculations. By adding the ability to convert between them, users get the best of both. The Awkward-RDataFrame bridge provides users with more flexibility in mixing different packages and languages in their analyses. It is a part of Awkward version 2. The implementation is being adopted by RootInteractive project~\cite{rootinteractive}. The user feedback is essential for further Awkward-RDataFrame development that is user-driven.

\section{Acknowledgment}
This work is supported by NSF cooperative agreement OAC-1836650 (IRIS-HEP) and NSF cooperative agreement PHY-2121686 (US-CMS LHC Ops).

\end{document}